\documentclass[sigplan,screen]{acmart}

\begin{document}

\title{A Micro-Service based Approach for Constructing Distributed Storage System}


\author{Yuhao Lu}
\email{luyuhao17@mails.ucas.edu.cn}
\affiliation{%
  \institution{ICT, CAS}
\streetaddress{ }
  \city{ }
  \state{ }
  \country{ }
  \postcode{}
}

\author{Zhenqing Liu}
\email{liuzhenqing@ict.ac.cn}
\affiliation{%
  \institution{ICT, CAS}
\streetaddress{ }
  \city{ }
  \state{ }
  \country{ }
  \postcode{}
}

\author{Dejun Jiang}
\email{jiangdejun@ict.ac.cn}
\affiliation{%
  \institution{ICT, CAS}
\streetaddress{ }
  \city{ }
  \state{ }
  \country{ }
  \postcode{}
}

\author{Liuying Ma}
\email{maliuying@ict.ac.cn}
\affiliation{%
  \institution{ICT, CAS}
\streetaddress{ }
  \city{ }
  \state{ }
  \country{ }
  \postcode{}
}

\author{Jin Xiong}
\email{xiongjin@ict.ac.cn}
\affiliation{%
  \institution{ICT, CAS}
\streetaddress{ }
  \city{ }
  \state{ }
  \country{ }
  \postcode{}
}

\renewcommand{\shortauthors}{Yuhao Lu, et al.}

\begin{abstract}
  This paper presents an approach for constructing distributed storage system based on
  micro-service architecture. By building storage functionalities using micro services, we can
  provide new capabilities to a distributed storage system in a flexible way. We take erasure coding
  and compression as two case studies to show how to build a micro-service based distributed storage
  system. We also show that by building erasure coding and compression as micro-services, the
  distributed storage system still achieves reasonable performance compared to the monolithic one.

\end{abstract}




\maketitle

\section{Introduction}
As data volume increases, interactive and data-intensive network services increase. Network services usually 
rely on large-scale distributed storage systems to achieve data access. In order to ensure user experience, 
especially the performance requirements of tail latency, a single IO request needs to distribute requests 
across thousands of servers so that it will generate too much load and should not be affected by other work.

In addition to the main IO storage function, the distributed system will also provide a variety of 
storage-related functions for users to use. For example, Ceph\cite{Ceph}, a widely used distributed storage system, 
provides users with compression and erasure coding functions in the RBD client and OSD server\cite{Rados_10.1145/1374596.1374606} modules respectively\cite{Ceph_268804}.
However, as the storage-related functions gradually increase, the monolithic architecture, the construction 
mechanism of these functions, brings negative problems to the main storage system. At the deployment level, 
multiple functions and IO storage are all executed in the same container, so that the peak load of the node 
will reach an extremely high level. At the program level, multiple functions and IO requests are carried out 
simultaneously in the same single program, which is bound to bring unexpected effects among various modules.

Our goal is to provide a better functional construction architecture for developers to develop distributed 
storage systems. In this case, we present a function construction mechanism from the perspective of code and 
function called microservice functional modules, based on the concept of microservices. Developer splits each 
module function into an independent service constructed by the microservice function module mechanism, and encapsulates 
the related code into a microservice module, which can be independently deployed and run on a different node 
to make itself compatible with IO application.

We apply the microservice architecture which serves from the front end of the network to the storage backend\cite{Microservices}, 
bringing the unique advantages of microservice to the developers. In this paper, we focus on Ceph\cite{Ceph}, 
and reconstructed its functional modules (compression, erasure) with microservice functional module 
construction mechanism. While maintaining its module relationship and functions, the storage engine 
execution model of its system functions is changed to reduce the pressure of single server and optimize 
the Ceph code structure to reduce the coupling between modules.

\section{System Design}

The microservice function module construction mechanism is designed to reconstruct the function by spliting it 
into an independent service, and encapsulate the relevant code into a microservice module, which can be 
independently deployed and run on a different node, so that it is separated from IO application and other 
modules. When IO application uses this function, it needs to call the microservice module through network 
communication instead of directly calling the exact code.

The microservice function module construction mechanism is designed to reconstruct the function, 
and split it into an independent service, and encapsulate the relevant code into a microservice module, 
which can be independently deployed and run on a different node, so that it is separated from IO application 
and other modules. When IO application uses this function, it needs to call the microservice module through network communication instead of directly calling the exact code.

\begin{figure}[htbp]
  \centering
  \includegraphics[width=1\linewidth]{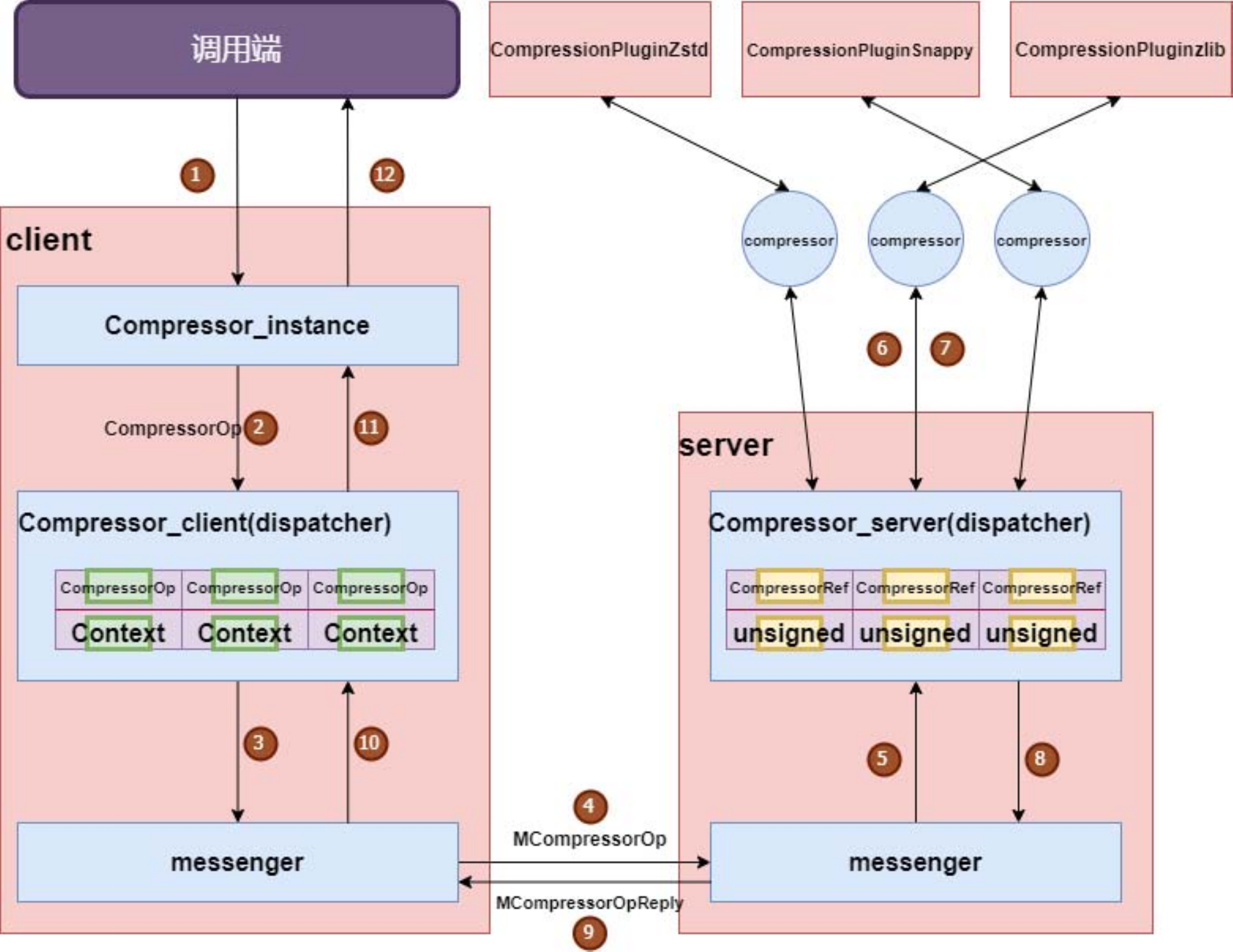}
  \caption{The structure and overall process of compressed microservices}
  \label{fig:compression}
\end{figure}

\begin{figure}[htbp]
  \centering
  \includegraphics[width=1\linewidth]{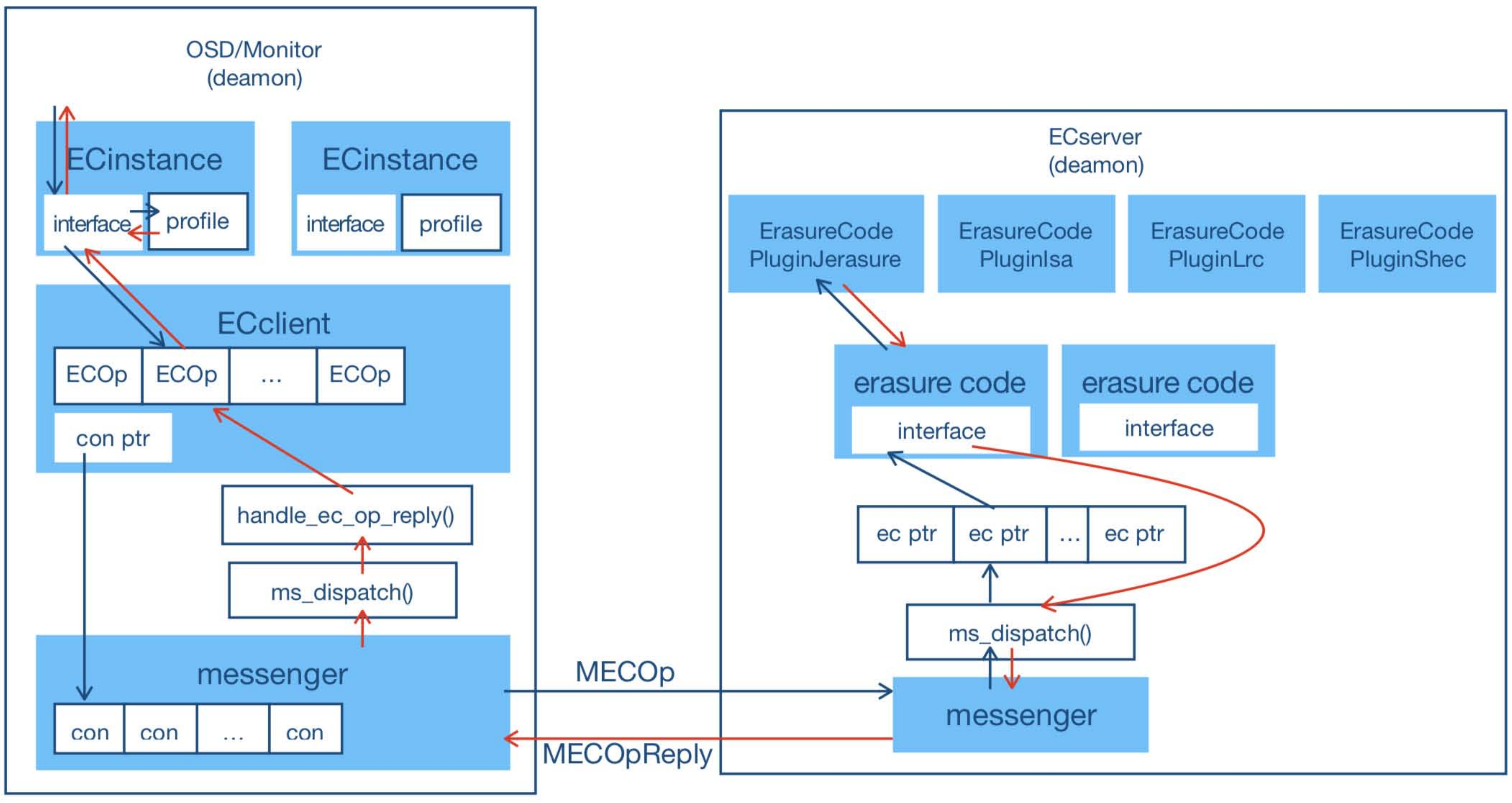}
  \caption{The structure and overall process of EC microservices}
  \label{fig:erasure}
\end{figure}

Figure \ref{fig:compression},\ref{fig:erasure} shows the process when user calls the exact functions with 
microservice function module construction mechanism. As it calls to start a related operation functions, 
a client panel, called Instance will be created to pack the function type and parameters into a request 
and save it into the queue of the Client. The Client then parses and encapsulates the compression request 
into a network message package and sends it to the Server through the Client’s messenger. As the Server 
receives the message it will unpack the packet and forward it to the corresponding function class for 
related operations. After the operation is done, the Server will encapsulate the data into a  packet, 
and send back to the Client. The Client then decapsulate the message and finally forward it back to user 
through the Instance.

When user uses the related function, the module keeps its original design is built on an independent node 
after receiving the message from the Server. More importantly, Network only needs to transfer the operation 
type and related parameters, while the rest of the extra overhead can be provided by the independent node.

\section{Evaluation}
\begin{figure}[htbp]
  \centering
  \includegraphics[width=1\linewidth]{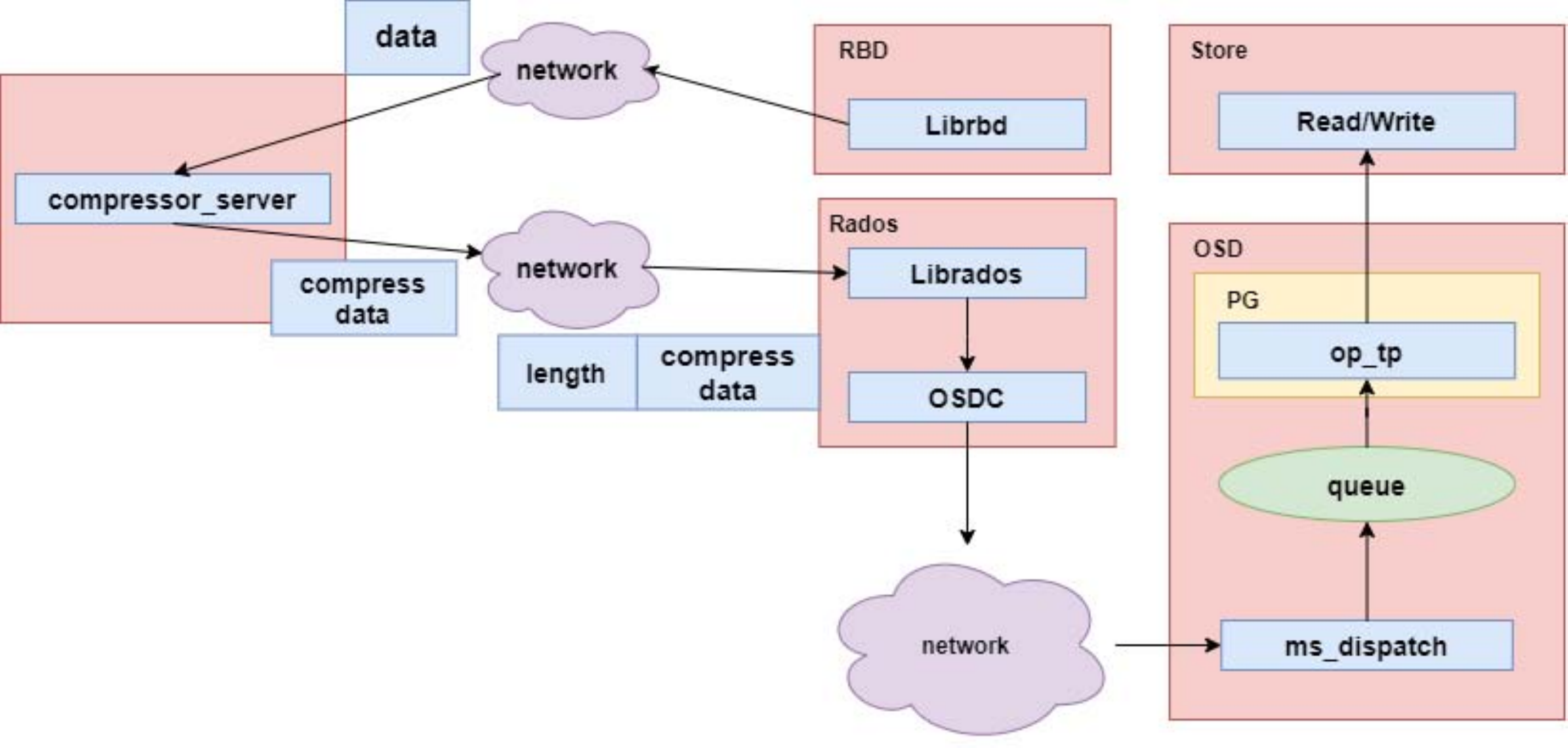}
  \caption{Compression function applied into IO application}
  \label{fig:compressor_instance}
\end{figure}

In this section, we implement two case studies on Ceph, one is compression,the other is erasure coding. 
We use two Huawei 2285 servers to build the original Ceph-12.2.5\cite{Ceph} system and the Ceph system using erasure 
correction and compression microservices. The environment configuration is shown in Table 
\ref{tab:TestConfiguration}. For erasure coding, we use Ceph's own IO performance evaluation tool, 
rados bench, to evaluate the IO bandwidth and latency of erasure objects in the original Ceph system 
and the system using erasure microservices. For compression, we use Fio to evaluate the performance of system.
\begin{table}
\centering
\caption{Test Configuration}
\label{tab:TestConfiguration}
\begin{tabular}{|c|c|}
\hline
Device model & Huawei2285 \\
\hline
CPU & (Intel(R) Xeon(R) CPU E5645@2.40GHz)*2 \\
\hline
Memory & 64GB \\
\hline
Network & async+posix \\
\hline
\end{tabular}
\end{table}
Figure \ref{fig:compressor_instance} shows how to merge the compression function into IO application. 
We insert it into the client rbd. When the client receives IO request, it will compress the data and send 
it to the OSD server\cite{Rados_10.1145/1374596.1374606} to store data. We use it to evaluate the IOPS, IO bandwidth and latency of compression 
in the original Ceph system and the system using compression microservices module. We evaluate the performance 
under different load and different chunk size(4KB,64KB) of data.

Table \ref{tab:Randwrite_4KB_IOPS} shows the performance of compression microservice (Instance) 
and original compression function (compressor) in Ceph IO randwrite mode with 4KB chunk size under 
different load. We find that under low load, IOPS and latency of microservices is slightly lower than 
the original function, while its is even higher than the original compression under high load. This 
shows that with small granularity, the performance of compressed microservices is close to that of the 
original version, and compression time is the main factor affecting performance, instead of network 
transmission time.
\begin{table}
\centering
\caption{Randwrite 4KB IOPS}
\label{tab:Randwrite_4KB_IOPS}
\begin{tabular}{|c|c|c|}
\hline
Iodepth & Instance & compressor \\
\hline
1 & 1000 & 657 \\
\hline
2 & 1680 & 1845 \\
\hline
4 & 2538 & 2836 \\
\hline
8 & 2827 & 2967 \\
\hline
16 & 2006 & 2709 \\
\hline
32 & 2601 & 2587 \\
\hline
64 & 2512 & 2443 \\
\hline
128 & 2457 & 2300 \\
\hline
\end{tabular}
\end{table}

Table \ref{tab:Write_64KB_BW} shows the performance of compression microservice (Instance) and original 
compression function (compressor) in Ceph IO write mode with 64KB chunk size under different load. 
We find that under low load, the bandwidth of the microservice is much lower than the original compression 
function, which is about 1/2 of its bandwidth, while under high load, as the queue depth increases, 
the bandwidth gap between them gradually decreases. When the load reaches 64 queue depth, the bandwidth 
of the microservice reaches its peak, which is about 0.75 of original. This shows that under large 
granularity and hign load, network transmission time is the main factor affecting performance. 
For reducing the pressure on a single server, optimizing the Ceph code structure, 
and decreasing the coupling between modules, the performance degradation is acceptable.
\begin{table}
\centering
\caption{Write 64KB BW(MB/s)}
\label{tab:Write_64KB_BW}
\begin{tabular}{|c|c|c|}
\hline
Iodepth & Instance & compressor \\
\hline
1 & 10.8 & 21.2 \\
\hline
2 & 18.7 & 37.5 \\
\hline
4 & 18.8 & 36.6 \\
\hline
8 & 19.8 & 35 \\
\hline
16 & 19.5 & 36 \\
\hline
32 & 19.1 & 36.2 \\
\hline
64 & 27 & 35.2 \\
\hline
128 & 24.2 & 35.5 \\
\hline
\end{tabular}
\end{table}

\section{Conclusion}

In this paper, we present a function construction mechanism from the perspective of code and function 
called microservice functional modules, based on the concept of microservices. We apply it to the storage backend, bringing the unique advantages of microservice to the developers. This mechanism reconstruct the function, and split it into an independent service, and encapsulate the relevant code into a microservice module, which can be independently deployed and run on a different node, so that it is separated from IO application and other modules. We focus on Ceph, and reconstructed its functional modules (compression, erasure) with microservice functional module construction mechanism. While maintaining its module relationship and functions, the storage engine execution model of its system functions is changed to reduce the pressure of single server and optimize the Ceph code structure to reduce the coupling between modules.

\bibliographystyle{ACM-Reference-Format}
  \bibliography{sample}
\end{document}